# Fabrication of functional 3D nanoarchitectures via atomic layer deposition on DNA origami crystals

Arthur Ermatov[1,†], Melisande Kost[2,†], Xin Yin[1], Paul Butler[3,4], Mihir Dass[1], Ian D. Sharp[3,4], Tim Liedl[1], Thomas Bein[2,*], Gregor Posnjak[1,*]

[1] Faculty of Physics and CeNS, Ludwig-Maximilians-Universität München, Germany

[2] Department of Chemistry and CeNS, Ludwig-Maximilians-Universität München, Germany

[3] Walter Schottky Institute, Technical University of Munich, Germany

[4] Physics Department, TUM School of Natural Sciences, Technical University of Munich, Germany

† These authors contributed equally to this project.

* Corresponding authors: Thomas Bein (thomas.bein@cup.lmu.de) and Gregor Posnjak (gregor.posnjak@physik.lmu.de).

## Abstract

While DNA origami is a powerful bottom-up fabrication technique, the physical and chemical stability of DNA nanostructures is generally limited to aqueous buffer conditions. Wet chemical silicification can stabilise these structures but does not add further functionality. Here, we demonstrate a versatile 3D nanofabrication technique to conformally coat micrometre-sized DNA origami crystals with functional metal oxides via atomic layer deposition (ALD). In addition to depositing homogenous and conformal nanometre-thin $ZnO$, $TiO_2$, and $IrO_2$ (multi)layers inside $SiO_2$-stablised crystals, we establish a method to directly coat bare DNA crystals with ALD layers while maintaining the crystal integrity, enabled by critical point drying and low ALD process temperatures. As a proof-of-concept application, we demonstrate electrocatalytic water oxidation using ALD $IrO_2$-coated DNA origami crystals, resulting in improved performance relative to planar films. Overall, our coating strategy establishes a tool set for designing custom-made 3D nanomaterials with precisely defined topologies and material compositions, combining the unique advantages of DNA origami and atomically controlled deposition of functional inorganic materials.

## Introduction

DNA origami[1] is a modular nucleotide-based self-assembly method offering molecular level tunability, thereby enabling access to a vast space of advanced nanostructures featuring diverse functions[2–11]. In the majority of previously demonstrated applications, DNA origami structures have been used in aqueous buffers since the stability of DNA double helices is highly dependent on pH, salt concentrations, and temperature[12]. To stabilise DNA origami structures, several approaches have been utilised, including covalent bonding[13], protein coating[14,15], and silicification using sol-gel chemistry[16–18]. However, the choice of materials that can overgrow DNA by virtue of wet chemistry is limited and these processes can be difficult to control. Moreover, the materials applied so far have focused on protecting the DNA nanostructure rather than adding further functionality. Coating precisely defined three-dimensional nanostructures with functional materials such as metals,[19] metal oxides, or heterojunctions could yield nanostructures that rival or even outperform 3D lithographic structures and open a path towards rationally designed 3D meta- and nanomaterials.

While silicification of DNA structures can greatly enhance their mechanical[18], chemical, and thermal stabilities[20,21], this wet chemical process is highly dependent on buffer conditions and is prone to poor reproducibility due to inhomogeneous coverage. Furthermore, significant effort is required to optimise the silica thickness and smoothness for the specific nanostructures to be coated. In addition, this approach introduces spatial design constraints due to the thickness of the silica shell, which cannot be controlled as precisely as other coating methods, such as atomic layer deposition (ALD). For intricate 3D structures with relatively small pores it would thus be highly beneficial to circumvent the silicification step and grow a broader range of functional materials with precise thickness control directly onto the DNA; however, capillary forces during drying can deform or collapse intricate nanoscale objects. While recent work[22] reported freeze drying of individual 3D DNA origami structures, this strategy requires uranyl formate incorporation for stability. Critical point drying (CPD) is an alternative method that reduces the deleterious effects of surface tension by replacing the solvent with a supercritical liquid and removing it in the supercritical phase. Thus, CPD may offer a route to drying bare DNA origami nanostructures while minimizing structural and morphological defects.

Atomic layer deposition  allows the growth of thin, conformal layers of materials, such as metal oxides or pure metals, using repeating cycles of self-limiting surface reactions, thereby enabling precise control of the material thickness down to the number of atomic layers[23]. Indeed, ALD functionalisation of DNA origami structures has been previously shown to be a powerful approach for coating substrate-supported monomeric structures (i.e., flat triangles and triangular prisms) with metal oxides such as aluminium oxide, titanium oxide, and hafnium oxide[24], as well as platinum[25]. In addition, vapour-phase chemical processes derived from ALD have been demonstrated to be effective in covering silicified 3D DNA nanostructures with metal oxides[26]. However, up to now, conventional ALD processes have not been demonstrated to fully penetrate complex 3D DNA origami structures, especially in the absence of $SiO_2$ stabilisation layers.

In this work, we coat micrometre-sized 3D DNA origami crystals with metal oxides using standard ALD processes, showing conformal coating and excellent penetration throughout crystals sized up to 10 μm. Not only do we apply this technique to $SiO_2$-stabilised crystal structures, but we used a low temperature ALD process to also coat bare DNA origami crystals prepared via a CPD process, thus demonstrating dry chemical stabilisation and functionalisation of complex 3D DNA nanostructures while preserving their geometry (Fig. 1). As a proof-of-concept application of this technique, we demonstrate the functional performance of $IrO_2$-coated DNA origami crystals deposited on fluorine-doped tin oxide (FTO)/glass supports for the oxygen evolution reaction (OER) via electrocatalytic water oxidation in aqueous environments.

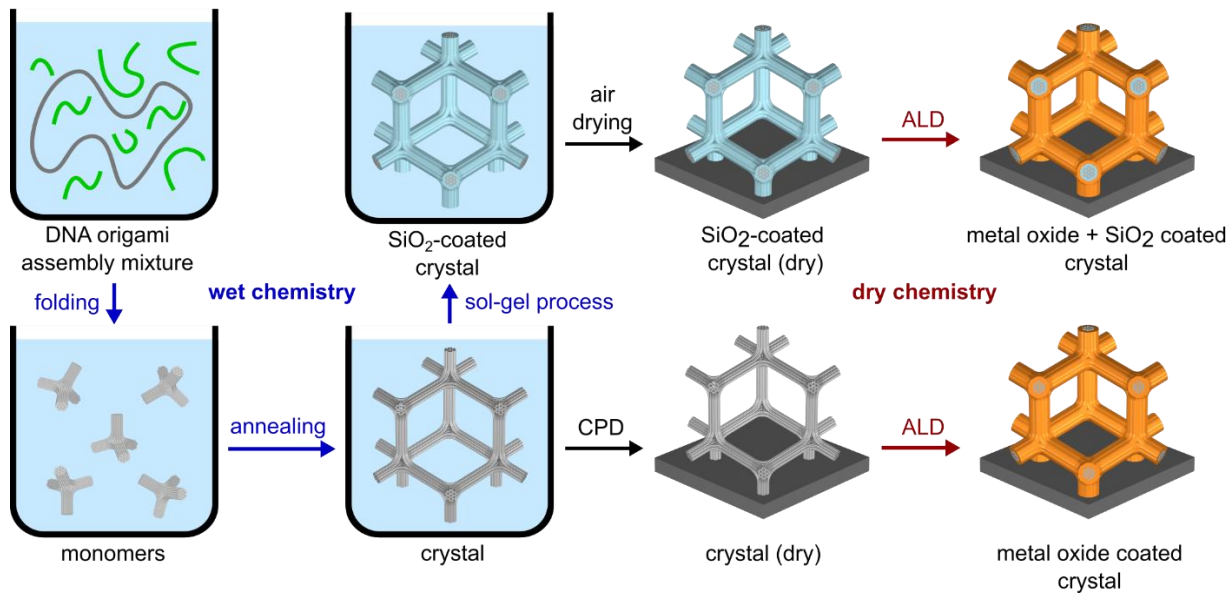

**Figure 1 | Overview of sample preparation.** Folding of DNA origami monomers, crystal annealing, and silica growth are performed in an aqueous 1x Tris-EDTA buffer solution comprising scaffold strand (gray), staple strands (brown), and magnesium ions (not shown). The silicified and bare crystals are then transferred to surfaces (glass or silicon) in liquid and dried in air (following silicification) or via CPD (bare structures without silicification). Finally, various functional metal oxide coatings are grown via ALD.

## ALD growth on crystals stabilised via sol-gel silicification

As the basis for our functional 3D DNA-based nanomaterials, we used DNA origami tetrapods that are designed to multimerise into a rod-connected diamond cubic crystal structure[27]. The resulting crystals are 5 – 10 μm in diameter and possess pore sizes of approximately 100 nm. The cross-section of the connecting rods consists of a 24-helix bundle with a diameter of 15 nm. To grow a layer of $SiO_2$ on the DNA origami with a sol-gel process, we first incubated the crystals in aqueous buffer with N-trimethoxysilyl-propyl-N,N,N-trimethylammonium chloride (TMAPS) before adding tetraethyl orthosilicate (TEOS) and acetic acid[16,27] (experimental details are provided in the Methods section). The silicification process was halted by washing the crystal with water and isopropanol, after which the crystals were concentrated and deposited onto silicon or glass surfaces by air-drying a droplet of the liquid containing the washed crystals. The silicification mechanically stabilises the DNA origami crystals, allowing air drying without deformation of the structures[20,27,28]. Subsequently, metal oxides were grown on the dried crystals using ALD. As a proof-of-concept, we used thermal ALD processes based on metal precursors including diethylzinc (DEZ), tetrakis(dimethylamino)titanium (TDMAT), titanium tetraisopropoxide (TTIP), and iridium acetylacetonate (Ir(acac)₃) to grow thin layers of ZnO, $TiO_2$, and $IrO_2$ with various thicknesses. Given the thermal stability of $SiO_2$-stabilised crystals, standard ALD recipes with reactor temperatures in the range of 175-250 °C were used (see Methods section for experimental details).

We first analysed the external surfaces of the ALD-coated crystals with scanning electron microscopy (SEM). In addition, to characterize the interior morphologies of the crystals, we used focused ion beam (FIB) milling to cut open the structures and performed cross-sectional imaging via SEM. The elemental compositions of the inner region of the crystal were subsequently determined using energy dispersive X-ray spectroscopy (EDX).

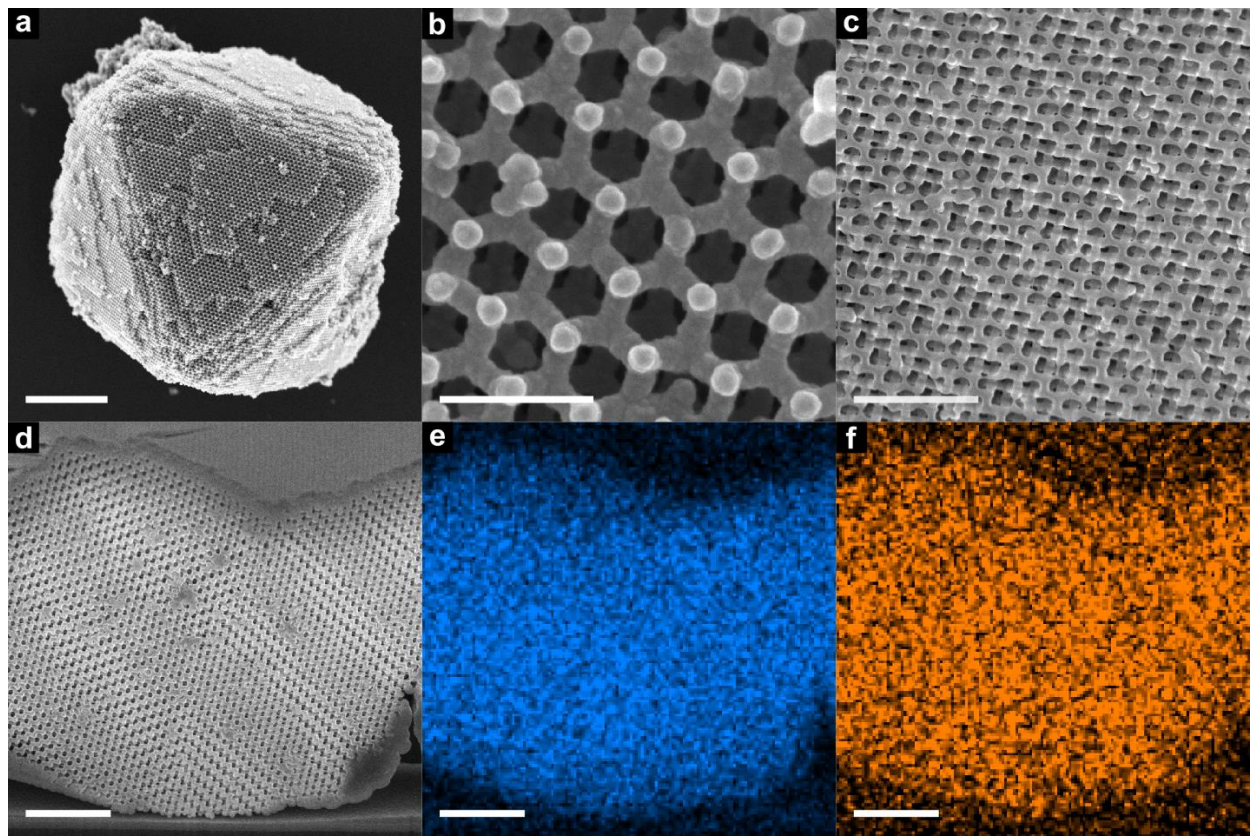

**Figure 2 | SEM images of DNA origami crystals coated with metal oxides. a,** Top view of an air-dried, silicified crystal after conformal coating with 10 nm of $TiO_2$ using ALD, scale bar 2 μm. **b,** Higher magnification image of the {111} surface of the crystal shown in (**a**), scale bar 200 nm. **c,** Image of the inner region of a crystal obtained by FIB milling of the crystal, scale bar 500 nm. **d-f,** Image and composition maps obtained following FIB milling of a silicified crystal coated with two consecutive shells of 5 nm $TiO_2$ (TTIP process) and 5 nm ZnO; SEM image (**d**) and corresponding EDX signals for Zn (**e**) and Ti (**f**); all scale bars are 1 μm.

Figures 2a-c show the monocrystalline morphology of the structures after ALD of $TiO_2$. The crystals primarily grew in octahedral shapes (Fig. 2a), with each of the eight triangular surfaces consisting of {111} planes of the diamond cubic lattice and exhibiting a regular hexagonal pattern (Fig. 2b)[27]. Remarkably, the outer surfaces of the crystals feature only very few and localised defects. Figure 2c shows the inner structure of the crystal that was exposed by performing a FIB cut perpendicular to the top crystal surface. The plane of the cut was not aligned with any of the crystal planes, resulting in a moiré pattern of interference fringes, where the periodic variation in thickness (Fig. 2c) originates from the different orientations of the cross-section through the local structure of the lattice.

Importantly, ALD results in conformal growth of the metal oxide throughout the interior of the crystal, as evidenced by the homogeneous distribution of deposited material observed by EDX mapping. In particular, there is no discernible gradient in the thickness of the coating from the centre towards the outer regions, as would be expected in the case of inhomogeneous infiltration of the ALD precursors. Furthermore, we do not observe selective filling or clogging of the lattice pores near the outer surfaces of the monocrystal. While localised aggregation and uneven thicknesses near the external surfaces of the crystal are observed in Fig. 2c, their absence in Fig. 2a indicates that these imperfections are a consequence of the FIB milling process. Moreover, if clogging of the pores would have occurred during ALD, it would have prevented further penetration into the structure, resulting in thinner coatings inside the crystal, which is not observed.

Many nanotechnological applications require heterostructures comprising several layers of different materials, each serving different purposes (e.g., passivation, charge separation and transport, photon management, chemical or catalytic function, etc.)[29,30]. To demonstrate the versatility and robustness of our sample preparation method for production of mesoscale nanolaminate architectures, we therefore grew heterostructures consisting of two different materials (ZnO and $TiO_2$, 5 nm thickness each) on top of each other (Fig. 2d-f). As for the case of a single ALD layer, we observe a homogeneous pore size (Fig. 2d) and elemental distribution (ZnO in Fig. 2e and $TiO_2$ in Fig. 2f) across the complete cross-section, indicating that the coverage of such heterostructures is comparable to that of a single material (Fig. 2a-c). Thus, this approach of combining both wet chemical and ALD processes enables the creation of mesostructured nanomaterials with smooth and homogeneous layers, in the present case using three different materials ($SiO_2$, ZnO, $TiO_2$). While this provides a versatile route to a vast range of complex nanomaterials, we now turn to investigating whether the wet chemical silicification step can be eliminated to grow ALD layers directly on DNA origami crystals.

## ALD growth on bare DNA origami crystals

In order to apply ALD processes on bare crystals, DNA origami structures must first be transferred from the liquid to the dry phase. However, air-drying of bare DNA crystals leads to severe morphological deformations and collapse of the intricate crystalline structures (Fig. S1). We attribute this instability to mechanical forces generated by surface tension as the liquid-gas interface moves through the crystal during drying. To overcome this problem, we used CPD based on liquid carbon dioxide ($CO_2$), which allows transition from the liquid to the gaseous phase via a supercritical fluid phase, thereby avoiding the propagation of a surface tension wave through the sample and associated drying artefacts. In our assembly protocol, bare DNA origami crystals are prepared in an aqueous folding buffer. Since water is poorly soluble in both liquid[31] and super-critical $CO_2$[32], the bare crystals were first washed with isopropanol, an intermediate solvent in which both water and $CO_2$ are soluble. A liquid droplet containing the crystals was then deposited on a ~ 1 cm² rectangle cut from a silicon wafer and placed in an isopropanol bath within the critical point dryer. Next, multiple purging steps were performed, during which liquid $CO_2$ diluted and replaced the isopropanol. Subsequently, the $CO_2$ was brought to the supercritical phase and removed (for the complete critical point drying process see Methods section). After drying, the crystals were coated with a 10 nm thick layer of $TiO_2$ via ALD using the TDMAT process (see Methods section).

While the CPD-dried DNA crystals do not show signs of degradation over the course of six weeks at room temperature (Fig. S2), one of the major obstacles to ALD growth on 3D DNA nanostructures is the potential degradation and deformation of the nanostructure at typical ALD process temperatures[33]. Such instabilities are particularly pronounced if the DNA structures do not have a protective layer, such as $SiO_2$, that increases their mechanical and thermal stability. Indeed, initial attempts to apply standard 200 °C ALD processes resulted in considerable deformation of the bare DNA origami structures (Fig. S3). However, by lowering the chamber temperature to 100 °C we were able to successfully deposit $TiO_2$ on bare DNA crystals while remaining within the ALD window and preserving their structural integrity (Fig. 3).

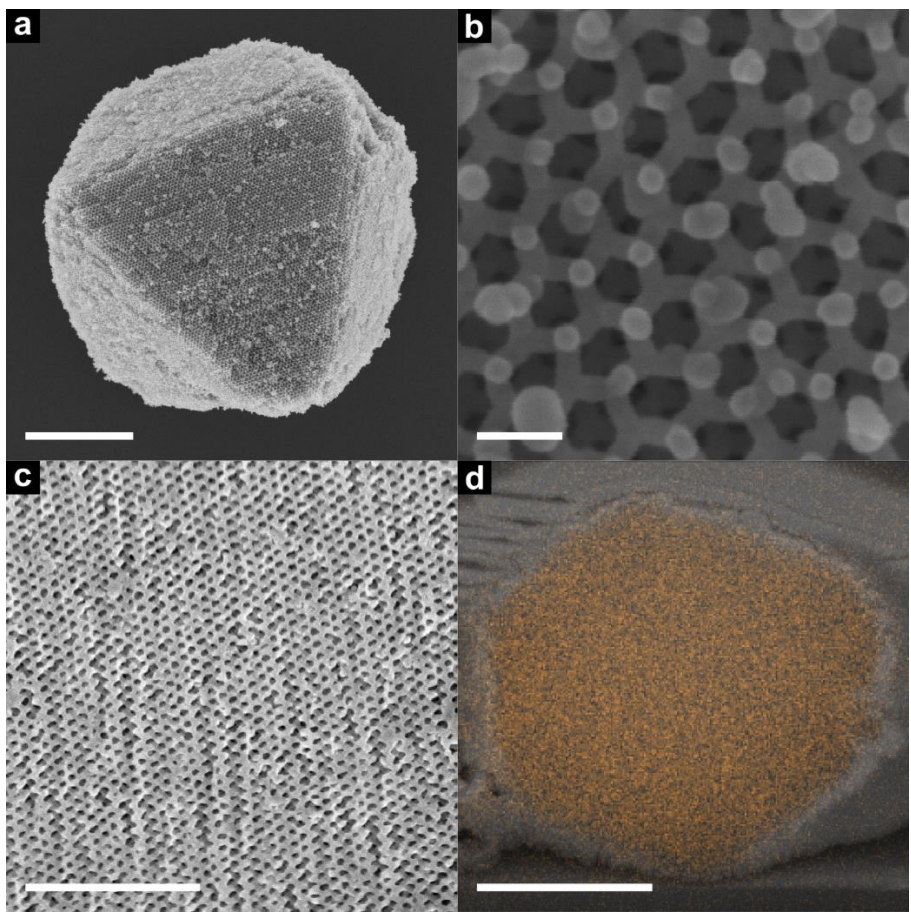

**Figure 3 | SEM images of CPD-dried crystals. a,** Top view of a DNA origami crystal dried using CPD and subsequently coated with 10 nm $TiO_2$ using ALD (TDMAT process, scale bar 2.5 μm). **b,** Close-up view of the crystal shown in (**a**) (scale bar 100 nm). **c,** Cross-sectional FIB cut of a CPD-dried crystal (scale bar 1 μm). **d,** Zoomed-out cross-sectional view of a FIB cut crystal, with the Ti EDX signal overlaid in orange (scale bar 2.5 μm).

The crystal morphology was investigated using SEM imaging of the crystal surface (Fig. 3a,b). In contrast to the air-dried bare DNA crystals, which appear deformed and collapsed (Fig. S1), the CPD-dried and $TiO_2$-coated DNA crystals retain distinct octahedral shapes (Fig. 3a), similar to the $SiO_2$-stabilised crystals (Fig. 2a-d). While the top surface of the crystal still displays a hexagonal pattern (Fig. 3b), individual hexagons are slightly distorted and less regular than the ones on the surface of the $SiO_2$-coated crystals in Fig. 2a,b. Furthermore, individual tetrahedron arms appear to be slightly bent. These distortions can be attributed to drying artefacts due to the reduced rigidity of the bare DNA structure compared to the silicified crystals. Nevertheless, long range ordering is retained and future optimisation of drying protocols may be applied to minimize these comparatively minor structural distortions.

Analogous to the $SiO_2$-stabilised crystals, the ALD-coated bare crystals were cut using FIB milling and both their structure and material composition were analysed using SEM and EDX, respectively. Remarkably, the overall crystal morphology (Fig. 3c) is preserved during the drying and ALD processes. However, the regular moiré patterns that were obtained for $SiO_2$-stabilised crystals (Fig. 2c-d) are not observed. We attribute this lack of interference between the cut surface and the crystal planes to the slight distortion of the crystal resulting from the drying process. Nevertheless, the quality of the deposition of $TiO_2$ on the bare DNA crystals is comparable to that obtained with the ALD process on $SiO_2$-stabilised crystals (Fig. 3d). Similar to the ALD of two different materials shown for $SiO_2$-coated crystals in Fig. 2, we expect this process to work equally for several layers of different materials, as long as the first material is deposited at chamber temperatures that do not compromise the mechanical stability of the DNA crystal.

## $IrO_2$-facilitated electrocatalytic water splitting

Among many other applications, the ALD-covered DNA origami crystals are potentially interesting for electrochemical,[34] photoelectrochemical, and photochemical[35] energy conversion due to the large surface-to-volume ratios associated with their porous structures. Furthermore, these crystals could form hierarchical porous systems with diverse topologies and a vast range of scaffold structures, compositions, and surface functionalities, all of which can be tuned to, for example, beneficially promote light absorption, charge carrier transport, mass transport, and catalytic activity. Here, we illustrate these opportunities with a proof-of-concept OER catalyst for electrochemical water splitting based on $IrO_2$-coated DNA origami crystals. Key questions that can be systematically addressed – and ultimately optimised – with DNA origami crystals include the electrical contact with a conducting electrode surface, electrical conductivity through the coated origami structure (depending on bulk oxide and Ir-oxide conductivity), catalytic surface area, mechanical stability of crystal attachment, chemical stability of the Ir-oxide phase, and oxygen gas evolution management. The latter will be strongly dependent on current density, pore structure, and layer thickness or crystal size. While a systematic study of these important aspects is beyond the scope of this work and will be the subject of future research, here we show that the Ir-oxide coated DNA crystals indeed feature significant electrocatalytic OER activity.

To generate electrochemically active anodes, we coated the DNA origami crystals with a ~3 nm thick layer of $IrO_2$[36] and used them as a catalyst for the OER[36] (Fig. 4). Using DNA origami monomers as the basis for the crystal, geometry and pore sizes can be precisely controlled. In this proof-of-concept demonstration, our crystals possess a pore diameter of approximately 100 nm and exhibit a surface-to-volume ratio of $1.88 \times 10^7$ m$^{-1}$ (see SI for calculation). These dimensions place our system in an intermediate regime compared to previous OER studies that have investigated the impact of pore sizes, either in the range of a few nanometres[37] or up to micrometres[38] in diameter.

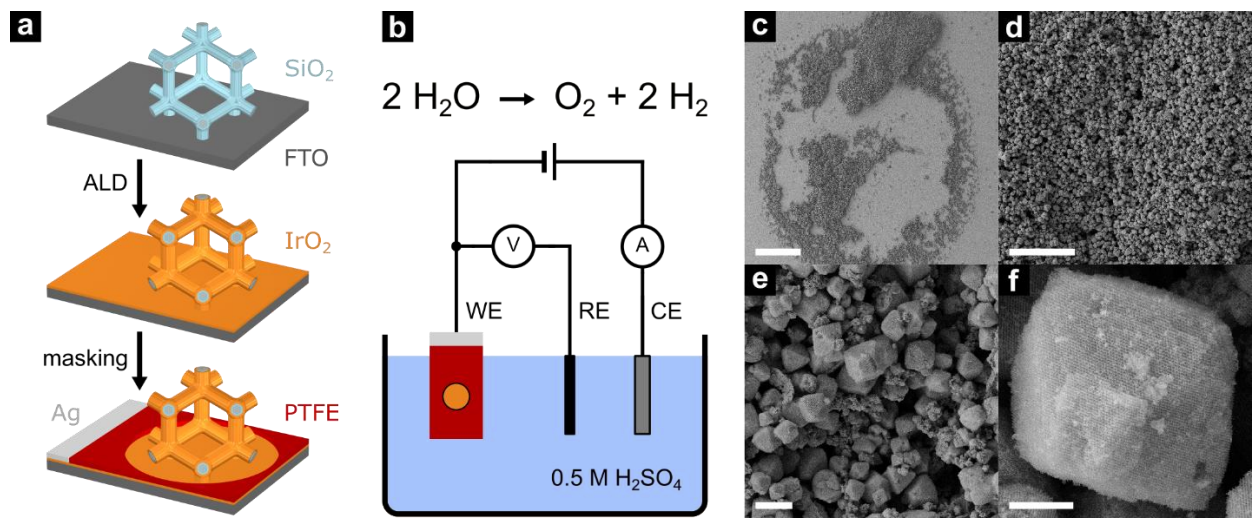

**Figure 4 | Experimental setup for electrocatalytic water oxidation using functionalised DNA origami crystals. a,** Sample preparation: Silicified DNA origami crystals were deposited and air-dried on an FTO-coated glass substrate. A thin layer of $IrO_2$ was subsequently grown via ALD, resulting in conformal deposition on both the exposed FTO substrate and on the $SiO_2$-coated DNA crystals. The substrate was then masked with PTFE, leaving an exposed 5 mm diameter circular area containing all of the DNA crystals. A separate exposed region on the edge of the sample was also defined, on which a Ag contact was placed. **b,** Experimental three-electrode setup for electrocatalytic water oxidation with $IrO_2$-coated DNA origami crystals. **c-f,** Exemplary SEM images of a 3x concentrated DNA origami crystal sample supported on FTO and coated with $IrO_2$ via ALD at different magnifications ((**c**) scale bar 500 μm, (**d**) scale bar 100 μm, (**e**) scale bar 10 μm, (**f**) scale bar 2 μm).

Silicified crystals were deposited on glass substrates coated with FTO (Fig. 4a,c-f). Since silicified DNA crystals can be air-dried, a solution containing the crystals can be applied to substrates and air-dried several times in a row, allowing for a degree of control over surface coverage and sample thickness. To determine the impact of surface loading, we examined two different coverages denoted as "1x DNA" and "3x DNA", where "1x DNA" comprises one deposition cycle with an estimated 0.28 pmol of crystal monomers, resulting in a total $IrO_2$-covered surface of approximately 20 cm$^2$, with the corresponding values for "3x DNA" being three-fold higher (note that these values are estimates based on the origami geometry, see SI for assumptions and calculation). FTO-coated glass slides without any deposited crystals were used as nominally planar reference samples. After drying, samples were coated with $IrO_2$ via ALD and were subsequently placed in a three-electrode electrochemical cell containing sulfuric acid (Fig. 4b). Cyclic voltammetry was performed with the $IrO_2$-coated samples configured as the working electrode and a fixed scan rate of 20 mV/s (Fig. 5a,b). While the first cycle was performed from 0 V vs. reversible hydrogen electrode (RHE) to 1.8 V vs. RHE, all subsequent cycles were run from 1.0 V vs. RHE to 1.8 V vs. RHE, for a total of 20 cycles. In the following, we discuss the key performance metrics obtained from these proof-of-concept DNA-origami based electrocatalysts, as well as future opportunities and needs for rational design and improvement of such catalytic systems.

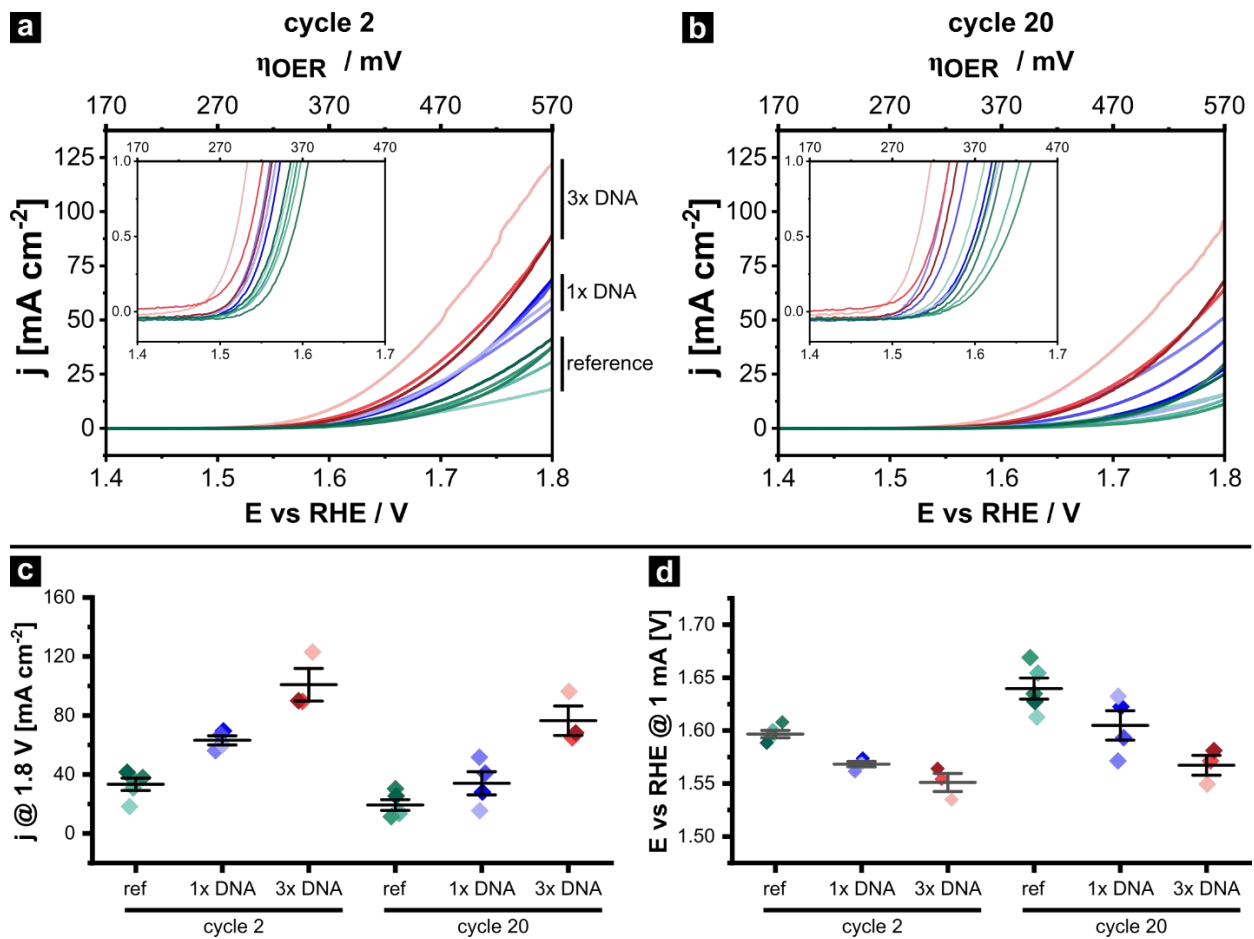

**Figure 5 | Electrolysis of water with IrO$_2$-coated DNA origami crystals.** The specific colour shade of each curve and point represents an individual sample. The reference ("ref") data (green shades) refer to samples that were prepared without adding DNA origami crystals to the FTO-covered glass. "1x DNA" (blue shades) denotes 0.28 pmol of crystal monomers deposited on the FTO-glass surface, "3x DNA" (red shades) denotes samples produced with three times as much material, as described in the text. **a,b,** Current density as a function of applied electrochemical potential measured in (**a**) cycle 2 and (**b**) cycle 20. Insets show zoomed-in regions near the onset potential. **c,** Statistical analysis of current density at 1.8 V vs. RHE. **d,** Statistical analysis of the electrochemical potential required to reach a current density of 1 mA/cm$^2$.

An approximately linear enhancement in catalytic activity with increasing crystal concentration is observed and is compared to the reference samples (Fig. 5c). However, the enhancement is not proportional to the increase in the surface area of the sample; although the "1x DNA" samples had an approximately 100x larger surface area than the reference sample, the activity enhancement with respect to the reference sample is only a factor of 1.9 (during cycle 2). For the "3x DNA" samples, the enhancement factor during cycle 2 was approximately 3 (for cycle 20: corresponding enhancement factors were 1.8 for "1x DNA", 4 for "3x DNA"). Nevertheless, we observe a decrease in the onset potential at 1 mA/cm$^2$ (Fig. 5d) with increasing density of crystals, reflecting the increase in the overall electrocatalytic activity. The narrow scatter range of the samples in Fig. 5d suggests a consistent IrO$_2$ coverage over the catalytically active surface.

The moderate current density increases suggests that only a fraction of the IrO$_2$ contributes to the overall activity, which likely originates from two factors that can be addressed in future work. First, while IrO$_2$ thin films possess high electrical conductivities[41] and the IrO$_2$ coverage inside individual crystals is homogeneous (Fig. S6), contact between crystals and to the substrate may limit electrical transport to active sites. To overcome this limitation, possible future strategies would include the integration of additional conducting layers prior to IrO$_2$ coating or directed growth of oriented DNA crystal films on (pre-functionalised) substrates. Second, nanoscale mass transport limitations arising, for example, from entrapment of evolved oxygen deeper within the crystals can reduce the contact area of water with the catalyst and hence limiting the reaction rate within the crystal. Previous work suggested that efficient detachment, transport, and coalescence of gas bubbles is also modulated by the 3D geometry of the catalyst, demonstrated by different catalytic behaviour for pore sizes ranging from 200 nm to 1200 nm[38]. This hypothesis can be tested in future experiments by using DNA crystal designs with different unit cell dimensions and topologies optimized for mass transfer – enabled by the versatile topological and geometric design of DNA crystals.

We further observed a decrease in catalytic activity after 20 CV cycles. We hypothesise that this is due to two reasons. First, IrO$_2$ is known to slowly dissolve under OER conditions[39,40], which is also observed in the present work and manifests for the reference samples as a decrease of the current density and increase of the overpotential between cycle 2 and cycle 20, as shown in Fig. 5. Second, crystals can detach from the surface (Fig. S4 (blue box) and Fig. S5a) due to mechanical stresses generated during bubble formation and evolution. Here, we note that the crystals are only physically deposited on the surface and adhere due to a combination of weak van der Waals forces and the thin metal oxide layer deposited via ALD, making them prone to mechanical detachment from the sample surface. This is further exacerbated by the "open" configuration of the working electrode electrochemical half-cell, which allows exfoliated crystals to disperse into the bulk electrolyte. By comparison, such effects can likely be suppressed in more confined designs, such as membrane electrode assemblies, or through the use of ionomer binders. Interestingly, the crystals not only appear to detach but also to partially re-attach to the IrO$_2$ surface at different locations (Fig. S4, red boxes), highlighting the dynamic nature of the working interface. Nevertheless, the morphology of the crystals themselves is not altered during the electrochemical experiment (Fig. S5b-d), indicating significant structural stability under harsh acidic conditions and demonstrating the robustness of the SiO$_2$-and ALD-coated crystals.

## Conclusion

We demonstrated the fabrication of 3D nanomaterials with controlled pore topologies and dimensions by coating micrometre-sized $SiO_2$-coated DNA origami crystals with metal oxides such as ZnO, $TiO_2$ and $IrO_2$. The porous 3D materials were conformally coated by ALD, with homogeneous layers extending through the complete inter regions of these intricate nanostructures. Furthermore, we demonstrated a method to transfer bare DNA crystals from the liquid to the dry phase while preserving their structures, allowing us to use ALD to grow $TiO_2$ directly on DNA. As an example application of the developed methods, we show a proof-of-principle study of $IrO_2$-based electrocatalytic water splitting, demonstrating enhanced catalytic performance compared to planar surfaces. The techniques demonstrated here greatly expand the range of applications based on DNA origami-based nano- and meso-structures, providing unique possibilities in terms of versatile design, precise molecular arrangement, and ALD-based functionalisation.

## Methods

### Folding and purification of DNA origami tetrapods

The DNA origami tetrapods[27] are designed based on a single 8634 nt long ssDNA scaffold strand and consist of four 35 nm long legs with a 24-helix bundle cross-section, corresponding to a 15 nm diameter. The folding mixture was prepared by mixing the scaffold strand (15 nM) with approximately 200 short staples (Integrated DNA Technologies, HPLC purified, 200 µM each in water, folding concentration 120 nM) in 100 µL aliquots containing 1x TE (10 nM Tris-HCl and 1 mM EDTA, pH adjusted to 8.0 with NaOH) and 20 mM $MgCl_2$. The temperature annealing protocol used for folding is listed in Table S1.

Table S1. Folding temperature protocol for tetrapod monomer

| Temperature (°C) | Time per °C (min) |
| --- | --- |
| 67 | 15 |
| 60-50 | 40 |
| 25 | storage |

The excess staples were removed by PEG precipitation after folding. The DNA origami solution was mixed with PEG solution (15 wt% PEG, 500 mM NaCl, and 1x TE) in a 1:1 ratio. When the volume of the DNA origami solution was less than 1 mL, it was diluted with 1x TE and 20 mM $MgCl_2$ buffer to 1 mL. This mixture was then centrifuged at 20,000 rcf for 30 min at 4 °C. A pellet was obtained by removing all of the supernatant from the tube. The pellet was re-dissolved in 1x TE and 5 mM $MgCl_2$ buffer by shaking at 800 rpm and 34 °C for 1 h. The concentration of monomers was then measured using a Nanodrop ND-1000 UV-Vis spectrometer and adjusted to 50 nM with 1x TE and 5 mM $MgCl_2$ buffer.

### Crystal growth

The diamond lattice DNA origami crystals were grown from purified tetrapod monomers (20 nM) mixed with the binding staples (200 nM) and the "mini scaffold" strands (8 µM) in 70 µL aliquots with 1x TE and 26 mM $MgCl_2$. A temperature algorithm was applied to the solution: first, a temperature ramp from 55 °C to 33 °C was performed at a rate of 2.5 h/°C, followed by incubation at 40 °C for 8 h. Then, 45 µL of supernatant was removed, and a second ramp from 55 °C to 33 °C (at a rate of -2.5 h/°C) was applied.

For CPD drying to obtain bare structures, the crystals were diluted to 2 mL with isopropanol, mixed, and left to sediment for 12 h, at which point the supernatant was carefully removed. This process was then repeated a second time. The crystals were then immediately processed (see Preparation of CPD-dried bare DNA crystal samples).

## Silicification

The remaining 20 µL crystal solution was mixed with 110 µL 1x TE and 5 mM $MgCl_2$ buffer in a 2 mL Eppendorf tube (final monomer concentration approximately 10 nM). After cooling to 4 °C, 1.6 µL of TMAPS (trimethyl[3-(trimethoxysilyl)propyl]ammonium chloride, TCI Germany, 50 % in methanol) diluted in methanol with a 1:3 ratio was added to the solution while shaking. After 30 min, 1.0 µL of TEOS (tetraethyl orthosilicate, Merck) was added to the solution while shaking. After 1 h, 1.3 µL of 0.5 M acetic acid was added while shaking. The temperature was then gradually increased, as summarised in Table S2. The silicified crystal was washed twice with water and twice with isopropanol by diluting the mixture to 2 mL, mixing, waiting at least 12 h for the crystals to sediment, and removing the supernatant.

Table S2. Temperature protocol for silicification

| Temperature (°C) | Shaking speed (rpm) | Time per °C (h) |
| --- | --- | --- |
| 4 | 800 | 4.5 |
| 10 | 800 | 2 |
| 16 | 800 | 2 |
| 22 | 800 | 16 |

## Preparation of air-dried SiO2-stabilised crystals samples

For $SiO_2$-stabilised crystal deposition, 1 x 1 cm$^2$ Si/$SiO_2$ chips with a 100 nm thermal oxide layer (MicroChemicals GmbH) were used as substrates. The chips were first washed in acetone for 30 s, then washed in isopropanol for 30 s, and then blow-dried with nitrogen. Afterwards, the chips were cleaned using a PICO plasma cleaner (Diener Electronic GmbH & Co. KG), utilising an oxygen plasma (45 sccm, 80% power) for 10 min.

After substrate preparation, the $SiO_2$-stabilised crystal solution (~20 µl) was pipetted onto the substrate and left to air-dry. Following drying, the substrate and supported crystals were coated via ALD according to the protocols described below.

## Preparation of CPD-dried bare DNA crystals samples

Bare DNA crystals kept in isopropanol were used immediately after their preparation. The same substrate preparation method was used as for the air-dried $SiO_2$-stabilised crystals (see above). Approximately 20 µl of bare crystal solution was pipetted onto each substrate and immediately transferred to a K850 Critical Point Drier (Quorum Technologies) chamber. The chamber was then immediately filled with isopropanol, not allowing the isopropanol-covered substrates to dry, and sealed. In all following steps, liquid levels were maintained to ensure full wetting of the substrates at all times until reaching supercriticality. The chamber was first cooled to 18 °C and liquid carbon dioxide ($CO_2$) was added to the chamber to dissolve and mix in the isopropanol until saturation. The system was then further cooled to 11 °C, where the remaining isopropanol was purged by flushing the chamber with liquid $CO_2$ for 10 min by fully opening the $CO_2$ inlet valve while adjusting the draining valve so that the liquid level in the chamber was kept constant. After 10 min of purging, both the drainage and the $CO_2$ valves were closed and the system was left for approximately 3 min to stabilise the $CO_2$ level and temperature at 11 °C. During stabilisation, the $CO_2$ inlet valve was opened as needed to adjust the $CO_2$ level in the chamber. The purging step was performed three times in total. After the last purging step, the chamber was fully filled with liquid $CO_2$ and cooling was turned off. The chamber was then slowly heated to 35 °C, surpassing the critical point of $CO_2$ (7.39 MPa, 31.05 °C) and thus transitioning from the liquid to the supercritical phase. After reaching 35 °C, a bleed valve was opened, which slowly reduced the chamber pressure to atmospheric pressure. When atmospheric pressure was reached, heating was turned off and samples were retrieved from the chamber.

## Preparation of FTO glass slides for subsequent IrO2 coating and electrocatalysis

FTO glass (TEC 15, sheet resistance 12-14 $\Omega/\square$) was purchased from XOP GLASS. The glass was cut into 1 x 2 cm² pieces with a diamond glass cutter and the FTO surface cleaned according to the following protocol. In a first step, the glass pieces were separated by spacers to ensure that their surfaces were accessible to the cleaning fluid. Next, 2 mL of Hellmanex II and 100 mL of bi-distilled water were added to cover the entire glass surface, which was then treated in an ultrasonic bath for 15 min. The liquid was then discarded and the samples were washed with bi-distilled water until all Hellmanex II was removed. The glass pieces were then immersed in bi-distilled water and treated in an ultrasonic bath for 15 min, after which the liquid was discarded. In the final treatment step, the glass pieces were fully covered with 100% isopropanol and treated in an ultrasonic bath for 15 min. Subsequently, the glass pieces were dried with nitrogen gas and masked using PTFE tape with a circular 5 mm diameter opening. The masked glass slides were stored in a dust-free container until usage. Before usage, substrates were cleaned using a PICO plasma cleaner (Diener Electronic GmbH & Co. KG), utilising an oxygen plasma (45 sccm, 80% power) for 10 min.

Six $SiO_2$-coated DNA origami crystal aliquots à 70 µl were combined, resulting in an estimated number of monomers of 8.4 pmol (assuming negligible losses in the preparation process, see surface enhancement discussion in SI). The combined aliquots were left to rest for 12 h for sedimentation. Afterwards, the supernatant was removed, and the remaining liquid was fully distributed among 12 "1x DNA" and 6 "3x DNA" samples. This was done by pipetting 0.4 µl of the combined aliquots onto the centre of the exposed region of each masked glass piece and letting it air-dry, repeating this step twice more for the "3x DNA" samples. This process was then repeated until all liquid was consumed or until there was not enough liquid to cover all samples according to the process. Neglecting potential losses, this yielded an estimated tetrapod quantity of 0.28 pmol for "1x DNA" samples and 0.84 pmol for "3x DNA" samples. For the reference samples, no crystals were pipetted onto the masked FTO glass.

All samples were then transferred to the ALD chamber for $IrO_2$ coating (see below). Before the ALD process, the masks were removed from the FTO glass pieces and after the ALD process the masks were reapplied.

### Atomic layer deposition (ALD) of ZnO using diethylzinc (DEZ)

ALD deposition was performed using a Picosun R-200 Advanced system (Applied Materials Inc.). The ALD ZnO precursor diethylzinc (DEZ) (Sigma Aldrich, ≤100%) was acquired from Sigma Aldrich and used as received. ZnO ALD was carried out at 175 °C chamber temperature. The nitrogen carrier gas flow during DEZ pulses was 150 sccm for all lines. DEZ was vaporised from a stainless steel container at room temperature; for this reactor, a precursor vapour pressure between 2 and 10 mbar is typically required to achieve reliable deposition. One ALD cycle included two 0.1 s DEZ pulses followed by 14.9 s of purging with pure gas, alternating with a one 1 s $H_2O$ pulse and subsequent 14 s of purging. Furthermore, a stop-flow protocol was established in order to allow the precursor vapor pressure to build up in the reaction chamber. Within the DEZ pulse, the low-flow protocol started 3 s before the DEZ pulse, at which point the flow in all other lines was reduced to 40 sccm for 8 s. Within this timespan, the chamber evacuation valve was closed 1 s before the precursor pulse and was kept closed for a total of 6 s. For the $H_2O$ pulse, the duration of the low-flow protocol was 7.8 s, whereas the chamber evacuation valve was kept closed for 6.3 s. Here, implemented delay times, as well as the flow rates, were kept the same. Apart from the low-flow rate, the flow rates of DEZ and $H_2O$ were 150 sccm. Using such a recipe, 22 cycles were performed for coating DNA crystals, which resulted in a layer thickness of 4.9 nm, as determined by spectroscopic ellipsometry on Si(100) reference substrates.

### Atomic layer deposition (ALD) of TiO2 samples using titanium tetraisopropoxide (TTIP)

ALD deposition was performed using a Picosun R-200 Advanced system (Applied Materials Inc.). The ALD $TiO_2$ precursor titanium isopropoxide (TTIP, 98 %) was acquired from Strem Chemicals and used as received. $TiO_2$ ALD was carried out at a chamber temperature of 250 °C. The nitrogen carrier gas flow was 150 sccm for all lines. TTIP was vaporised from a stainless steel container at 85 °C. One ALD cycle included two 0.1 s TTIP pulses followed by 14.9 s of purging with pure gas, alternating with one 1 s $H_2O$ pulse and subsequent 14 s of purging. Furthermore, a stop-flow protocol was established in order to allow the precursor vapor pressure to build up in the reaction chamber. Within the TTIP pulse, the low-flow protocol started 3 s before the TTIP pulse, at which point the flow in all other lines was reduced to 40 sccm for 8 s. Within this timespan the chamber evacuation valve was closed 1 s before the precursor pulse and was kept closed for a total of 6 s. For the $H_2O$ pulse, the duration of the low-flow protocol was 7.8 s, whereas the chamber evacuation valve was kept closed for 6.3 s. Here, implemented delay times, as well as the flow rates, were kept the same. Using such a recipe, 160 cycles were used for coating DNA origami structures, which resulted in a layer thickness of 5.1 nm, determined by ex-situ spectroscopic ellipsometry on Si(100) reference substrates.

### Atomic layer deposition (ALD) of TiO2 samples using tetrakis(dimethylamino)titanium (TDMAT)

Titania films were also deposited by a thermal ALD process using a Fiji G2 system from Veeco Instruments Inc. For these depositions, tetrakis(dimethylamido)titanium (TDMAT, 99.999 % Sigma-Aldrich, heated to 75 °C) was used as the titanium precursor and water (filtered at 0.2 μm, VWR Chemicals) was used as the oxidant. The precursor and carrier gas lines were heated to 110 °C, while the chamber walls and substrate stage were heated to 100 °C (200°C for the sample in Fig. S3). In the first half-cycle, the TDMAT was pulsed into the chamber for 0.25 s followed by a 25 s purge; in the second half-cycle, water was pulsed into the chamber for 0.06 s, then the chamber is purged for another 25 s. During the deposition, the chamber was kept at a base pressure of 0.24 Torr, and argon was used as the carrier and purge gas with a flow rate of 420 sccm throughout the entire procedure. The thickness of the resulting titania film was monitored with in-situ spectroscopic ellipsometry measured on a silicon reference chip that was placed in the reactor together with the samples.

## Atomic layer deposition (ALD) of IrO2 samples using iridium acetylacetonate (Ir(acac)3)

ALD deposition was performed using a Picosun R-200 Advanced system (Applied Materials Inc.). ALD $IrO_2$ precursor $Ir(acac)_3$ (98 %) was acquired from Strem Chemicals and used as received. $IrO_2$ ALD was carried out at a chamber temperature of 188 °C. During pulses, the nitrogen carrier gas flow was 40 sccm for all lines. $Ir(acac)_3$ was sublimated from a borosilicate glass container at 186 °C. Ozone was supplied by an ozone generator (INUSA AC2025) from a feed of 500 sccm 1 %vol $N_2$ in $O_2$ (Air Liquide, 99.9995 %). Each cycle consisted of three $Ir(acac)_3$ pulses and one ozone pulse, separated by purge intervals. An $Ir(acac)_3$ half cycle comprised a 1.6 s pulse, 5 s static exposure, and 10 s purge time. For the ozone half cycle, the times were 4 s, 2 s and 10 s, respectively. A total of 80 cycles were used for the OER catalyst coatings, which resulted in a layer thickness of 3.1 nm, as determined by ex-situ spectroscopic ellipsometry on Si(100) reference substrates.

## Electrocatalytic experiments

Electrochemical characterisation was performed in a three-electrode setup in a quartz cell at room temperature. The cell was filled with 20 ml 0.5 M $H_2SO_4$ (Sigma-Aldrich, Titripur® volumetric standard) electrolyte. A PGSTAT302N potentiostat/galvanostat (Methrohm Autolab B.V.) equipped with a FRA32 M impedance analyser was connected to a hydroflex reversible hydrogen electrode (Gaskatel Gesellschaft für Gassysteme durch Katalyse und Elektrochemie mbH) or $Hg/HgSO_4/K_2SO_4$ (sat.) (REF601 radiometer analytical, Hach Company) reference electrode for cyclic voltammetry (CV) and galvanostatic (chronopotentiometry) measurements, respectively.

To avoid artefacts arising from the formation of trapped oxygen bubbles in porous electrode layers, as described in the work of El-Sayed et al.[42], cyclic voltammetry (CV) measurements were used to determine the electrochemical activity of the catalysts, which allows to reduce oxygen bubbles during each cathodic scan.

The CV measurements were performed with iR-correction (95%) in a potential window of 0.0 – 1.8 V versus RHE with a scan rate of 20 mV/s for the first scan and 1.0 – 1.8 V versus RHE and same scan rate for subsequent 19 scans. Before each CV measurement, impedance spectroscopy was performed at 0.5 V versus RHE to determine the corresponding electrolyte resistance from the high-frequency region ($R_s$).

## SEM and EDX imaging

SEM images in Fig. 2a-d, Fig. 3c,d, Fig. S3, S6 and S7 were recorded with an FEI Helios Nanolab G3 UC scanning electron microscope equipped with a field emission gun operated at 2 – 20 kV. EDX measurements were recorded at an operating voltage of 20 kV with an X

MaxN Silicon Drift Detector with an 80 mm$^2$ detector area (Oxford Instruments) and AZTec acquisition software (Oxford Instruments).

SEM images in all other figures were collected using the SEM feature of an eLine electron beam lithography tool (Raith GmbH) operated at 4 kV acceleration voltage, utilising the in-lens or secondary electron detector. Samples that were not coated using ALD (Fig. S1) were sputter coated prior to imaging using a S150 Sputter Coater (Edwards) with a 60:40 Au/Pd target for 30 s, thereby ensuring sufficient electrical conductivity to avoid charging-induced artifacts.

### Focused ion beam (FIB) milling
The area of interest for a cross section of the oriented DNA origami crystal was covered by a protective 1.5-2 µm thick Pt layer with an ion beam (Fig. S7). After depositing the protective coating, the DNA origami crystal was milled with the focused gallium ion beam to form a thin vertical slab. The acceleration energy of the ion beam was 30 kV throughout the process, with an initial beam current of 2.5 nA. The slab thickness was gradually reduced via ion milling until a final thickness of approximately 1-1.3 µm was reached. The beam current was decreased in conjunction with the reduction in the slab thickness to minimize beam damage in the area of interest: 0.77 nA was used down to a thickness of 2 µm, 0.4 nA was chosen in the range 2 - 1.5 µm, and a current of 0.25 nA was then used until the final thickness was reached.

## Acknowledgements

We would like to thank Dr. Steffen Schmidt for his help with the FIB lithography and the corresponding SEM and EDX images. M.K. would like to express her gratitude to Robert von Kalle for the reliable and secure transportation of the samples between the various locations. A.E., X.Y, M.D., T.L. and G.P. acknowledge funding from the ERC consolidator grant 'DNA Funs' (Project ID: 818635) and the Federal Ministry of Education and Research (BMBF) and the Free State of Bavaria under the Excellence Strategy of the Federal Government and the Länder through the ONE MUNICH Projects Munich Multiscale Biofabrication and Enabling Quantum Communication and Imaging Applications. T.B. gratefully acknowledges funding from the German Federal Ministry of Education and Research (BMBF) within the Kopernikus Project P2X: Flexible use of renewable resources – research, validation, and implementation of 'Power-to-X' concepts (project numbers 3SFK2P0-2 and 3SFK2Z0-2) as well as project 03HY129C IRIDIOS, and from the Bavarian research network "Solar Technologies Go Hybrid". The authors thank the e-conversion Cluster of Excellence (Deutsche Forschungsgemeinschaft (DFG, German Research Foundation) EXC 2089/1 – 390776260) for financial support.

## Competing interests
The authors declare no competing interest.

## Contributions

A.E., X.Y., and G.P. prepared the DNA origami crystals. A.E. prepared the dry DNA origami crystal samples. P.B. performed ALD coverage of CPD-dried samples with $TiO_2$ (TDMAT process). M.K. performed ALD coverage of $SiO_2$-stabilised samples with $IrO_2$, ZnO and $TiO_2$ (TTIP process). M.K. performed FIB lithography and electrocatalysis experiments. A.E. and M.K. performed SEM and EDX measurements, designed the electrocatalysis experiments, analysed and interpreted the resulting data. A.E., M.K. and M.D. conceived the idea and designed the ALD experiments. A.E. and G.P. wrote the manuscript with input from all authors. I. D. S., T. L., T.B. and G. P. guided the project.

# Supplementary Information

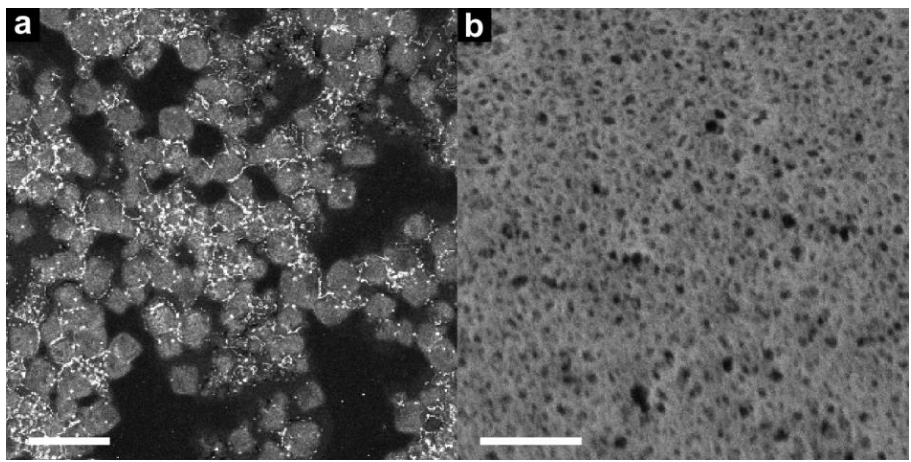

**Figure S1 | Bare crystals air-dried from buffer solution. a,** Plan-view SEM image of bare crystals air-dried from the buffer solution (crystals were coated by sputtering Au/Pt for SEM imaging, see Methods section), scale bar 20 μm. **b,** Zoomed-in image of an exemplary bare crystal shown in (**a**), scale bar 400 nm.

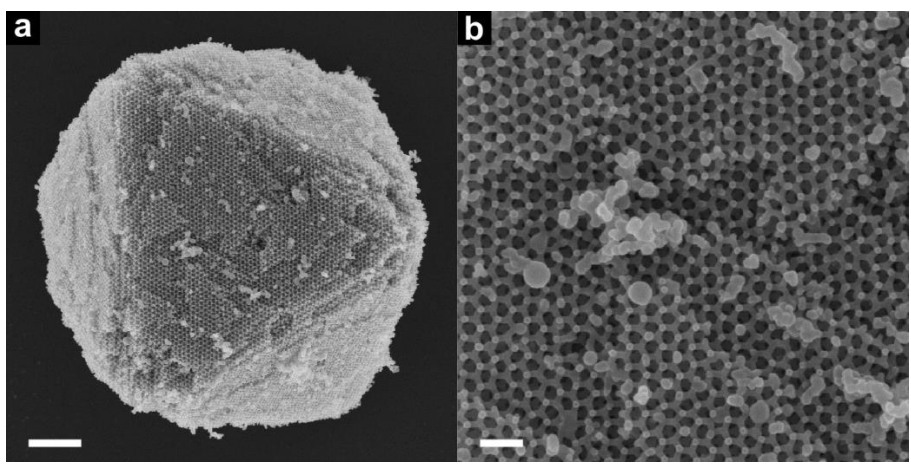

**Figure S2 | CPD-dried crystal stored under ambient conditions for six weeks and subsequent coating with 10 nm of TiO$_2$ using the TDMAT process. a,** scale bar 1 μm. **b,** Zoomed-in view of (**a**), scale bar 200 nm.

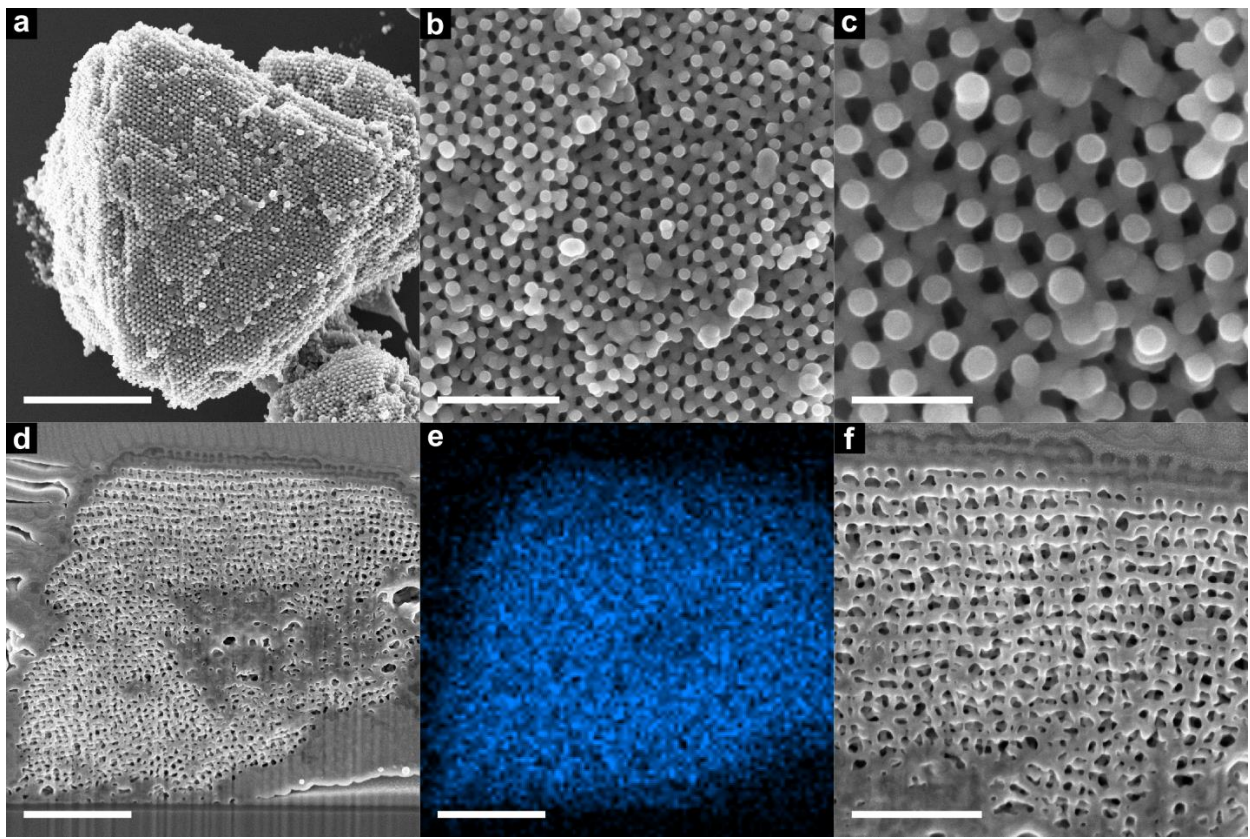

**Figure S3 | CPD-dried bare crystals coated in TiO₂ (TDMAT process) using a chamber temperature of 200 °C. a,** SEM image of an exemplary crystal, scale bar 2 μm. **b-c,** Zoomed-in images of (**a**), scale bar 500 nm (**b**) and 200 nm (**c**). **d,** SEM images of the interior of a CPD-dried, bare crystal coated in 10 nm TiO₂ using FIB milling, scale bar 1 μm. **e,** Ti EDX signal of the cut crystal shown in (**d**), scale bar 1 μm. **f,** Zoomed-in image of crystal shown in (**d**), scale bar 500 nm.

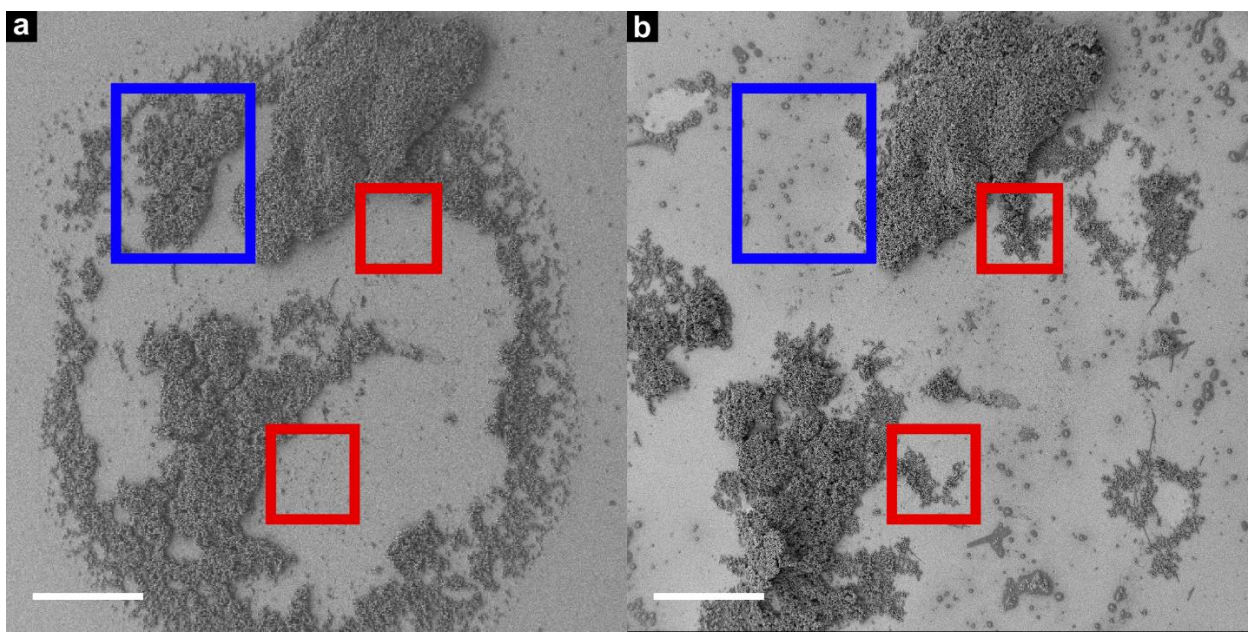

**Figure S4 | Comparison between "3x DNA" catalyst samples before and after electrocatalysis. a,** "3x DNA" sample before electrocatalysis (same as Fig. 4c), scale bar 500 nm. **b,** "3x DNA" sample after 20 cycles of electrocatalysis (same as Fig. S5a), scale bar 500 nm. Note the change in crystal distribution: in some regions, crystals detach during the experiment (exemplary region shown in blue box) and re-attach at different locations (exemplary regions shown in red boxes). Both scale bars 500 nm.

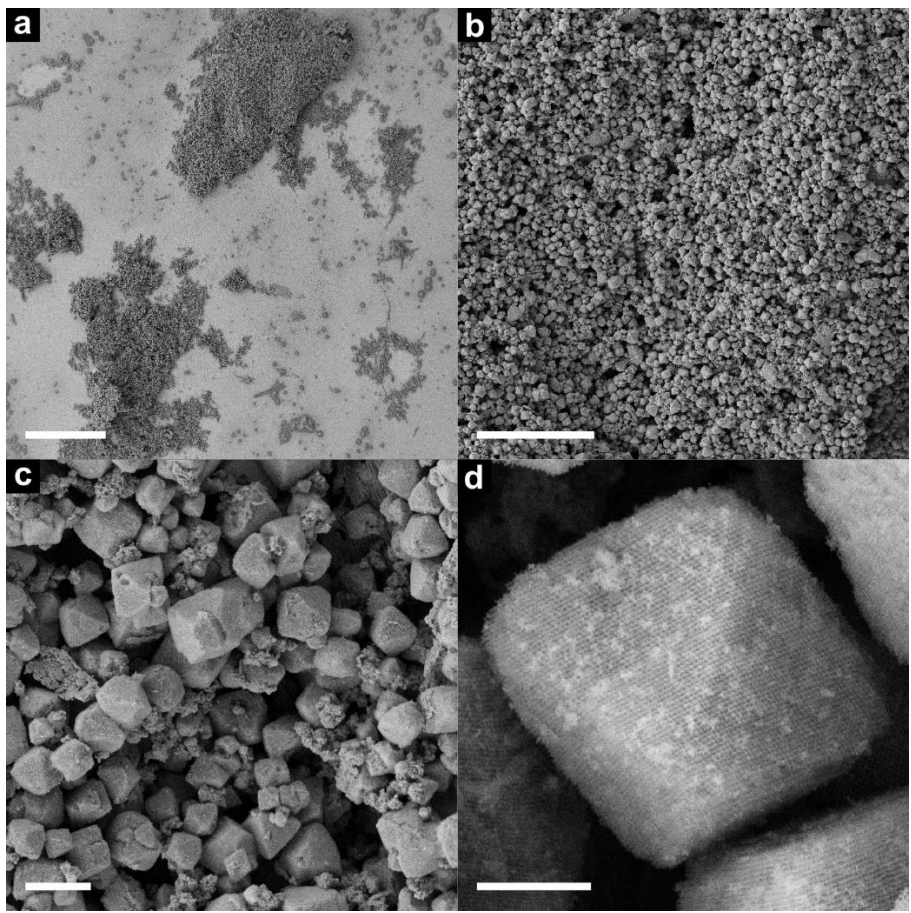

**Figure S5 | SEM images of "3x DNA" electrocatalysis sample of Fig. 4c-f after 20 cycles of electrocatalysis at different magnifications. a,** scale bar 500 μm. **b,** scale bar 100 μm. **c,** scale bar 10 μm. **d,** scale bar 2 μm.

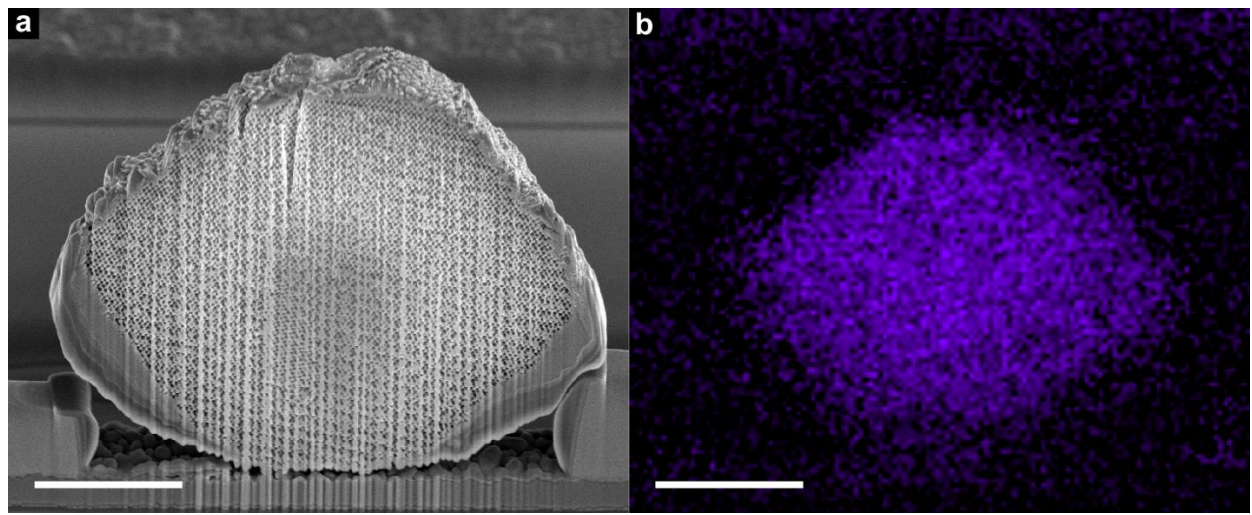

**Figure S6 | Silicised DNA origami crystal coated with ALD IrO₂. a,** Cross-sectional SEM images of the interior of an SiO₂-stabilised crystal coated in IrO₂, scale bar 2.5 µm. Note that the damage to the surface and the stripe-like artefacts are a result of the FIB milling process. **b,** Ir EDX signal of the cut crystal, scale bar 2.5 µm.

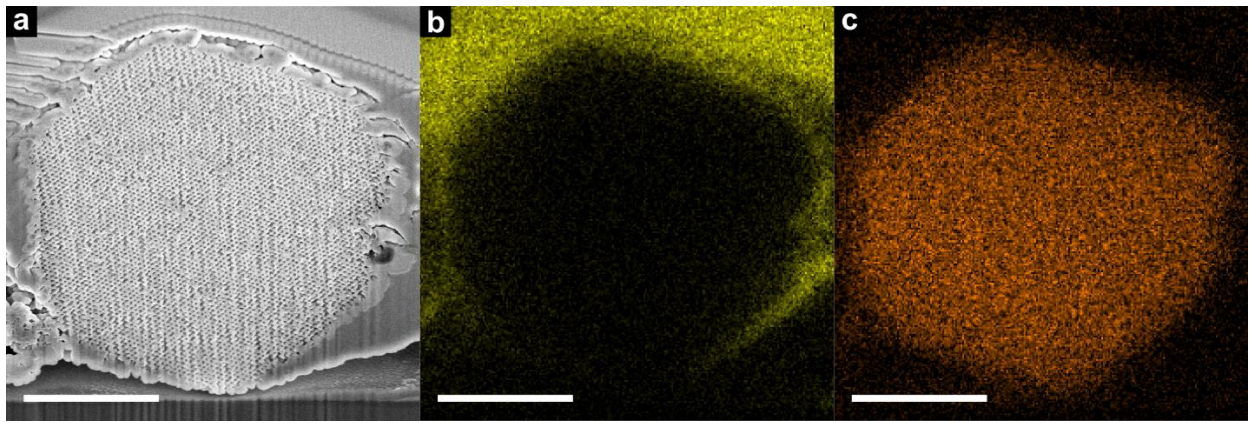

**Figure S7 | Crystal preparation using FIB milling. a,** SEM imaging of cut crystal. **b,** Pt EDX signal of of (**a**), depicting the Pt layer that protects the area of interest from gallium ion bombardment. Ion milling of the remaining crystal not covered in Pt yields a cross-section, allowing observation of the inner morphology of the DNA crystal shown in (**a**). **c,** Ti EDX signal of (**a**), showing the signal of interest inside the DNA crystal. All scale bars 2 µm.

## Estimation of surface enhancement due to DNA origami crystals

The calculation of the enlarged surface area by crystal deposition on the FTO glass slide is based on several assumptions:

- the folding yield of the DNA origami tetrapods is 100%,

- there are no losses in the tetrapod purification process,

- there are no losses in the crystal washing steps,

- all crystals remain in solution (i.e. do not stick to tube walls or pipette tips before deposition),

- all crystals in solution are distributed between the FTO glass substrates according to the pipetted volume, and

- no crystals are swept away due to external forces during the ALD process.

The following calculation therefore only provides approximate estimates for the order of magnitude of the surface area. From experience with the system, we estimate the errors to be around 20 % of the values obtained by the calculation.

One crystal aliquot à 70 µl contains 20 nM of scaffold. Assuming perfect yield, this results in

$$n_{\text{tetrapod}} = V_{\text{aliquot}} \cdot c_{\text{scaffold}} = 70 \, \mu l \cdot 20 \, \text{nM} = 1.4 \cdot 10^{-12} \cdot N_A = 1.4 \, \text{pmol}$$

per aliquot (for six combined aliqouts, the number of tetrapods is therefore 8.4 pmol).

The surface of every tetrapod can be approximated by four cylinders of length 35 nm, ignoring the circular surfaces that are covered by adjacent tetrapods. The radius is given by half the thickness of the DNA origami (~7.5 nm), the $SiO_2$ layer (~2.5 nm) and the $IrO_2$ layer (~ 3.1 nm), resulting in a total radius of 13.1 nm. The resulting surface area per aliquot is therefore

$$A_{\text{surface, aliquot}} = n_{\text{tetrapod}} \cdot 4 \cdot l_{\text{cylinder}} \cdot 2\pi \cdot \left( r_{\text{DNA}} + r_{\text{SiO}_2} + r_{\text{IrO}_2} \right)$$

$$\Rightarrow A_{\text{surface, aliquot}} = 1.4 \cdot 10^{-12} \cdot N_A \cdot 4 \cdot 35 \cdot 10^{-9}\,\text{m} \cdot 2\pi \cdot \left( (7.5 + 2.5 + 3.1) \cdot 10^{-9}\,\text{m} \right)$$

$$\Rightarrow A_{\text{surface, aliquot}} = 9.7 \cdot 10^{-3}\,\text{m}^2 = 97\,\text{cm}^2$$

Since 6 aliquots combined resulted in the equivalent of 30 "1x DNA" or 10 "3x DNA" samples, this means that, per sample, the surface is given by

$$A_{\text{1x DNA}} = \frac{6 \cdot A_{\text{surface, aliquot}}}{30} = 19 \cdot 10^{-4}\,\text{m}^2 \approx 20\,\text{cm}^2,$$

and

$$A_{\text{3x DNA}} = 3 \cdot A_{\text{1x DNA}} = 57 \cdot 10^{-4}\,\text{m}^2 \approx 60\,\text{cm}^2,$$

respectively.